# Spiketrum: An FPGA-based Implementation of a Neuromorphic Cochlea

MHD Anas Alsakkal, Jayawan Wijekoon, *Member, IEEE*

*Abstract*—This paper presents a novel FPGA-based neuromorphic cochlea, leveraging the general-purpose spike-coding algorithm, Spiketrum. The focus of this study is on the development and characterization of this cochlea model, which excels in transforming audio vibrations into biologically realistic auditory spike trains. These spike trains are designed to withstand neural fluctuations and spike losses while accurately encapsulating the spatial and precise temporal characteristics of audio, along with the intensity of incoming vibrations. Noteworthy features include the ability to generate real-time spike trains with minimal information loss and the capacity to reconstruct original signals. This fine-tuning capability allows users to optimize spike rates, achieving an optimal balance between output quality and power consumption. Furthermore, the integration of a feedback system into Spiketrum enables selective amplification of specific features while attenuating others, facilitating adaptive power consumption based on application requirements. The hardware implementation supports both spike-based and non-spike-based processors, making it versatile for various computing systems. The cochlea's ability to encode diverse sensory information, extending beyond sound waveforms, positions it as a promising sensory input for current and future spike-based intelligent computing systems, offering compact and real-time spike train generation.

*Index Term*—Spike-based coding; Neuromorphic Engineering; Cochlea; FPGA-based implementation; Gammatone Filters.

## I. Introduction

As the realm of neuromorphic engineering continues to flourish, the past few decades have been marked by a significant surge in the development of spike-based hardware. Researchers around the globe are increasingly invested in enhancing the computational capability of interactive devices [1-5], responding to an ever-growing demand for high-speed, efficient sensory data processing. A pivotal element in this pursuit is the real-time, efficient encoding of vast quantities of sensory information, including but not limited to audio signals. The process of encoding plays an indispensable role in enabling intelligent decision-making mechanisms by converting raw data into a format that can be easily interpreted and utilized by the system. Efficient encoding is crucial for preserving essential audio characteristics, ensuring precise interpretation and response to diverse sounds. This method yields numerous advantages, including reduced storage demands, improved data transfer speeds, and enhanced processing efficiency. Moreover, it effectively reduces audio data size, resulting in swifter and more accurate recognition.

The choice of representation used in this encoding process, especially for audio signals, can greatly impact the overall efficiency of the system. A preferred choice in neuromorphic engineering is spike-based representation, renowned for its speed and low power consumption [6]. The complexity of audio features with their complex time-frequency characteristics further underscores the need for sophisticated encoding schemes. The utilization of Gammatone kernels in sound feature extraction is a significant move in sound processing research. Inspired by the human auditory system, these filters have proven effective in various applications such as speech recognition, speaker recognition, and music classification. What sets Gammatone kernels apart is their unique ability to simultaneously capture temporal and spectral information within a sound signal, a quality vital for distinguishing between diverse sounds. Furthermore, these kernels exhibit remarkable noise robustness, making them an invaluable tool for extracting meaningful features even from noisy sound signals, a trait of utmost importance in real-world applications where sound signals are frequently tainted by environmental noise [7, 8].

Numerous electronic cochlea models have been proposed over the past three decades, each varying in complexity and abstraction levels [1, 4, 5, 9-15]. These models aspire to mimic some mechanisms exhibited by the biological ear, drawing insights from its remarkable auditory processing capabilities. However, a critical challenge lies in the balance between mimicking biological accuracy and ensuring efficient, low-power hardware implementations. Cochlear models are typically classified as either active or passive, with active models implementing recurrent feedback inputs. They can also be one-dimensional (1D) or two-dimensional (2D) models, with 1D models primarily mimicking the propagation of the hydraulic wave across the Basilar Membrane (BM) based on its longitude. These models divide the BM length into equal parts, representing the BM using auditory filterbanks or transmission lines (TL) [16]. Although TL models provide an accurate representation of wave propagation on the BM, their complex differential equations make them less hardware-friendly compared to filter-based models [17]. It's worth noting, however, that the use of 1D cascaded filters to mimic wave propagation on the BM, while appearing biologically plausible, is associated with several issues. These include noise accumulation as the signal propagates through numerous filters, system-wide breakdowns due to the failure of an intermediate filter, and significant output latency, which poses a challenge for real-time processing applications. Some of these problems have been addressed by introducing 2D cochlea models [18], which provide multiple paths for incoming signals to propagate between cascaded cochlear sections using resistive elements, or by using parallel filters [5, 19].

While closely imitating complex biological sensory architectures could lead to a better understanding of the techniques utilized by such systems [10, 20-22], their direct hardware implementation may not always result in efficient or low-power solutions. Hence, developing an advanced cochlea model at a higher level of abstraction with efficient spike-based encoding is desirable. This enables the design of neuromorphic cochlea models that leverage the latest digital electronic technologies while capitalizing on ultra-low-power implementations. Such an advancement leads to the creation of a superior sensor, addressing the evolving needs of both contemporary and future spike-based intelligent computing systems and neural implantable devices. The Field Programmable Gate Array (FPGA)-based configurable hardware is a well-suited choice for applications requiring adaptable configurations, primarily in research and development tasks, even though it may come at the expense of increased power consumption. Its flexibility allows for fine-tuning hardware to specific requirements, making it a valuable asset in dynamic and experimental contexts [3].

This paper examines a proposed hardware realization of the general-purpose spike-coding algorithm, Spiketrum [23]. Building upon the growing recognition of Gammatone filters as effective emulators of auditory nerve fibres [23], we have implemented Spiketrum using 40 rectangular band (ERB) Gammatone filters. To validate Spiketrum's functionality, the hardware implementation is rigorously compared with software counterparts, showcasing its competence in distinguishing different audio signal classes by feeding the output spikes to a neural network model. Our proposed neuromorphic cochlea introduces a robust and efficient coding paradigm tailored for neuromorphic hardware technologies, ideal for demanding tasks such as speech recognition and sound classification. This cochlea excels in real-time processing, offering adjustable output spike rates, thus increasing flexibility for the efficient encoding of complex audio signals. Furthermore, we introduce the concept of feedback projections to the algorithm, seeking to emulate the intelligent adaptability observed in biological cochlea. These adaptations have the potential to selectively amplify certain signal characteristics while attenuating others, contributing not only to computational efficiency but also significant power savings.

The following sections present the proposed architecture of the neuromorphic cochlea (Section II), details about the hardware techniques and the design considerations used by the FPGA-based implementation (Section III), experimental results followed by a sound-classification application (Section IV), discussion and comparisons to other prior works (Section V) and conclusion with final remarks (Section VI).

## II. SPIKETRUM: SYSTEM ARCHITECTURE

Spiketrum is designed to characterize and convert time-varying analogue signals, typically associated with auditory data, into highly efficient spatiotemporal spike patterns. It offers a sparse and efficient coding scheme, allowing precise control over spike rates. The system is based on three fundamental stages: Feature Extraction, Residual Computing, and Intensity-to-Place Coding, as can be seen in Fig. 1-a. Within the Feature Extraction stage, the incoming signal undergoes a transformation via a matching pursuit signal decomposition method [24], employing a predetermined dictionary comprising 40 Gammatone kernels. More on the process of generating the kernel set can be found in [23]. These kernels exhibit a remarkable resemblance to the frequency selectivity of the human auditory system. The selection of the most fitting kernel for the observed segment of the input signal is accomplished through convolution operations. This process yields sophisticated codes encapsulating spatial position denoted as '$m_i$', precise temporal positioning denoted as '$\tau_i$', and the convolution intensity denoted as '$s_i$', which characterizes the correspondence between the signal and the selected Gammatone kernel. It is worth emphasizing that each code $(m_i, \tau_i, s_i)$ functions as an effective repository of the unique attributes inherent to the input signal, ensuring the preservation of distinguished natural features.

The algorithm's iterative nature underscores its focus on prioritizing the encoding of significant features, followed by the processing of less pivotal ones. This entails the removal of the previously encoded kernel component from the current input segment before generating a new code. The Residual Computing unit plays this pivotal role in selectively eliminating the influence of already captured features, ensuring the representational distinctiveness of each code in relation to the auditory signal. This process unfolds in three stages (Shifter, Multiplier and Subtractor) as shown in the 'Kernel Elimination' part of Fig.1 -a. At the beginning of each subsequent iteration, the central position of the highest matching kernel $\emptyset_{m_i}(t)$ from the previous iteration is adjusted to its temporal positioning '$\tau_i$' aligning it temporally with the point of maximal correlation between the processed input audio segment and the selected kernel to produce $(\emptyset_{m_i}(t - \tau_i))$. Subsequently, this shifted kernel is scaled by the intensity of the captured code '$s_i$' to produce $(s_i \emptyset_{m_i}(t - \tau_i))$. The results is then subtracted from the processed segment '$x(t)$' producing the residual signal '$x_{new}(t)$' (i.e., $x_{new}(t) = x(t) - s_i \emptyset_{m_i}(t - \tau_i)$), which overwrites the Signal RAM, preparing for the next code generation operation.

The Spiketrum algorithm offers a notable advantage with minimal conversion errors in waveform-to-spike transformation. In specific low-power contexts like battery-operated robotics, a strategic compromise between encoding accuracy and resource efficiency emerges as a favorable option. Beyond power conservation, enhancing the algorithm involves integrating feedback projections to emulate adaptive qualities seen in biological cochlea. These adaptations empower the algorithm to selectively amplify certain features, refining captured signal characteristics. This configuration promises a more efficient encoding architecture, adept at tackling intricate audio processing challenges, like the cocktail party problem. Moreover, the generated codes play a pivotal role in the Feedback block, influencing when to conclude the encoding process.

In the Intensity-to-Place (ITP) coding stage, the generated codes undergo a mapping process, resulting in binary spike trains on output channels. This stage is carefully designed to minimize any loss of information during the transition from continuous waveforms to discrete spikes. Each kernel is associated with a certain number ($N$) of output channels (also referred to as output fibres) corresponds to a distinct intensity level or firing threshold.

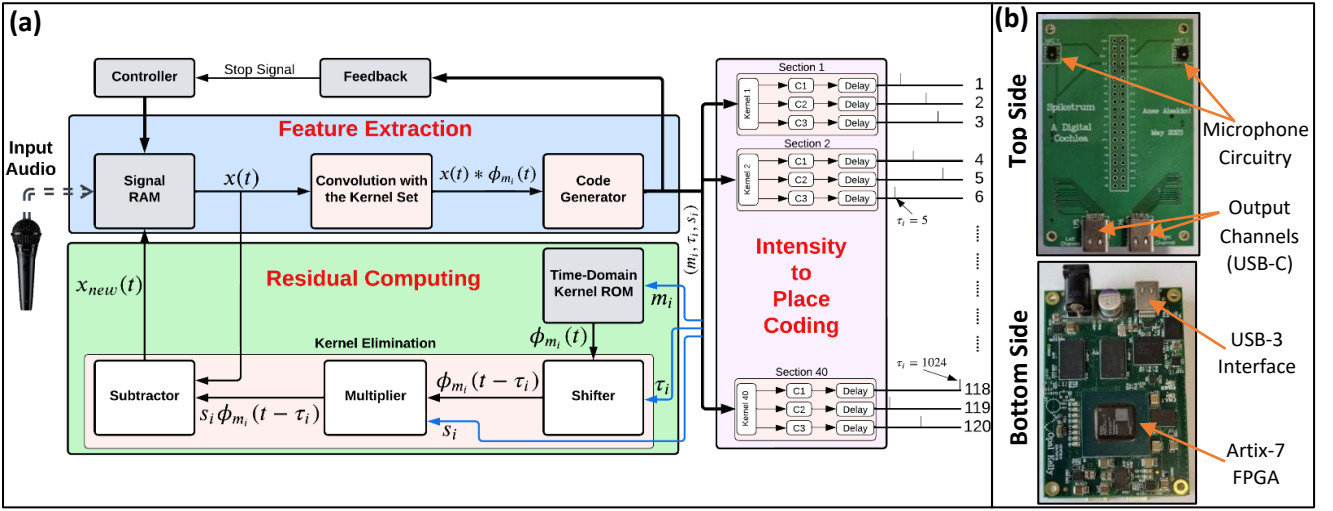

**Fig. 1:** showcases the hardware implementation of the proposed FPGA-based neuromorphic cochlea using the Spiketrum spike-encoding algorithm. In **(a)**, the hardware architecture is outlined, depicting the Feature Extraction, Residual Computing, and Intensity-to-Place Coding processes. **(b)** The real-time FPGA-based prototype is illustrated. It offers live audio streaming through two microphones or USB-3 interface and produces output spikes via 120 channels through a USB-C port.

Based on statistical analyses presented in [23], the optimal number of output channels per kernel, $N$, that affords high encoding sparsity while maintaining practical resource utilization, has been determined to be three. These three output channels are characterized by logarithmically distributed centre intensities, specifically denoted as $C_i$(0.0065, 0.4115, and 25.8744). Since each of the 40 implemented kernels has three associated output channels, Spiketrum has a total of 120 output channels (see Fig. 1-a).

When a code is generated by the Code Generator, the ITP block reads the code and select the section associated with the selected kernel, denoted as $m_i$. Subsequently, from the three channels linked to the matching kernel, the channel intensity, $C_i$, that most closely aligns with the code's intensity, $s_i$, is chosen. This matching process involves evaluating all three channels (C1, C2, and C3) associated with the $m_i$ kernel. After pinpointing the channel with nearest intensity, a time delay equivalent to $\tau_i$ is introduced by the Delay process, followed by the emission of a spike through the output channel associated with the code's kernel index, $m_i$, and the closest intensity level $C_i$. The Intensity-to-Place Coding efficiently preserves the spatiotemporal information inherent in the signals, facilitating the transformation of dense, continuous inputs into a discrete format while retaining the original signal's characteristics.

The generated spikes can be decoded with no loss of information, and the original waveform can be easily reconstructed, facilitating accurate calculation of the Spiketrum's encoding error. This reconstruction primarily involves the linear superposition of the small waveforms represented by the generated spikes. The temporal placement of each spike indicates the presence of its corresponding waveform in the time domain, while the channel number conveys information about the kernel index (small waveform) and intensity of the selected waveform. Thus, by using the Spiketrum output spike patterns, the original sound signals can be reconstructed using the following equation:

$$\hat{x}(t) = \sum_i c_i \phi_{m_i}(t - \tau_i) \qquad (1)$$

Where $\hat{x}(t)$ is the reconstructed signal, $c_i$ is the convolution intensity between the selected kernel $\phi_{m_i}$ and the input segment, and $\tau_i$ is the temporal position (time shift).

It is evident that the encoding error decreases as the spike rate increases [3, 23]. Additionally, the algorithm's feasibility for implementation in conventional FPGAs, known for their cost-effective resource utilization, enhances accessibility and scalability. This salient attribute underscores the efficacy and reliability of Spiketrum as a robust encoding framework for neural signal processing, holding promise across a spectrum of scientific and technological applications. It is essential to note that Spiketrum's applicability extends beyond the realm of audio signals, as it demonstrates remarkable versatility across diverse input domains.

### III. SPIKETRUM: DESIGN CONSIDERATIONS AND IMPLEMENTATION DETAILS

This section provides an in-depth exploration of the FPGA-based implementation of Spiketrum, emphasizing its design considerations and execution specifics. While the algorithm could leverage cutting-edge ultra-low-power Very Large-Scale Integration (VLSI) technologies for optimal power efficiency, our focus here is on the FPGA-based implementation. This decision aligns with the need for configurability, allowing users to finely adjust parameters like spike rates for a nuanced balance between output quality and power consumption. Beyond these considerations, factors such as the algorithm's real-time adaptability, enabling dynamic configurations in response to changing stimuli, and its seamless integration with various neuromorphic hardware architectures contribute to tailoring performance to specific application requirements. It's important to note that FPGA designs, with the desired configurations, can smoothly translate into ultra-low-power VLSI when a customized solution with additional power efficiency becomes necessary.

The proposed neuromorphic cochlea is implemented on XEM7310 Opal Kelly board that features Xilinx Artix-7 FPGA as shown in Fig. 1-b. XEM7310 is integrated with many useful peripherals such as SDRAMs, EEPROMs and clock generators. To facilitate real-time processing, segments of input audio signals are captured, amplified, and digitized through a specially designed microphone auxiliary circuit. These processed signals then relayed to the cochlea. Alternatively, pre-recorded audio signals can be streamed directly to the device through a high-speed USB-3 interface. The cochlea generated output spikes are made readily accessible through a USB-C port. This output port facilitates seamless integration of the cochlea with neuromorphic systems through direct hardware device connection. It also allows for the configuration of communication protocols, including Address-Event Representation (AER).

The proposed hardware adopts a 34-bit fixed-point architecture to achieve the desired computational precision with low power consumption, prioritizing resource efficiency [25]. In contrast, floating-point architectures prioritize computational accuracy. Our approach strikes a balance between power efficiency and precision by utilizing a high bit resolution within the fixed-point architecture, approaching the capabilities of floating-point solutions. The choice of a 34-bit resolution aligns with the maximum data sample precision supported by Fast Fourier Transform (FFT) cores and forms the foundation for the Spiketrum algorithm's hardware architecture. Subsequent subsections will delve into hardware details and techniques employed by individual units in the proposed implementation. To optimize resource utilization, on-chip memories feature a Native interface (chosen over AXI4 to reduce FPGA slices) and are generated using the Xilinx Block Memory Generator (BMG) IP Core with the Minimum Area Algorithm. This method minimizes the number of required block memory primitives. For additional details on these interfaces and techniques, refer to [26].

*A. Feature Extraction*

The adapted Matching Pursuit (MP) algorithm [24] is employed to decompose any signal into a linear combination of shifted and scaled waveforms from a standard set of templates (kernel set). The decomposition utilizes the Convolution operation to identify the most suitable kernel combination for the incoming audio signal segment. Two distinct hardware architecture approaches were explored for the Feature Extraction unit, each tailored to different optimization goals to meet diverse requirements in various applications.

The first strategy, termed the "area-optimized" approach, prioritizes conserving hardware resources and minimizing the overall implementation footprint. However, this comes with trade-offs in the quality of the encoded signal during real-time processing. In this configuration, convolution operations are executed in the time domain to determine the optimal matching kernels for the processed segment of the input signal. A single DSP slice is employed for multiplication and accumulation (MACC) operations, ensuring efficient resource utilization. Running at the maximum available clock speed of 200 MHz, this approach can achieve a spike rate of 16 spike per segment (sps), equivalent to 368 spikes per second, with real-time processing. The primary objective of the "area-optimized" approach is to create a compact implementation suitable for low-cost FPGA-based systems or integration into low silicon area integrated circuits (ICs). Such implementations prove advantageous in applications where considerations such as power consumption and hardware footprint are critical, for instance, in cochlear implants.

The second approach, termed "performance-optimized", prioritizes processing speed by leveraging additional resources to reduce computational load, making it ideal for FPGA implementation. This strategy excels in handling high computational demands, delivering a substantial performance boost suitable for real-time scenarios requiring high-quality input representation with elevated spike rates. Despite a 52% increase in Block RAMs (BRAMs), 31% more Digital Signal Processing (DSP) slices, and a 6% rise in Look-Up Tables (LUTs), the "performance-optimized" approach, with real-time processing, can deliver up to 80 sps (1,840 spikes per second), outperforming the "area-optimized" approach by 5 times at the same clock speed of 200 MHz. This makes it particularly well-suited for real-time, high-accuracy processing applications such as autonomous vehicles and brain-computer interfaces (BCIs). In terms of power consumption, the "performance-optimized" approach exhibits an approximate 75% increase in power usage compared to the "area-optimized" alternative. The choice between the "area-optimized" and the "performance-optimized" approaches hinges on the specific requirements of the application, taking into account factors such as required quality of input sound wave encoding, resource utilization, power consumption, and processing speed.

This work focuses on the FPGA-based implementation of a "performance-optimized" approach. We provide a detailed overview of hardware specifications and techniques employed. Incoming audio segments are continuously captured and stored in the single-port 8.7 kB Signal RAM (Fig. 1-a), which also serves for audio segments received from Residual Computing output. A custom controller oversees Signal RAM operations, managing read/write processes and coordinating data transfer to the FFT Core for subsequent multiplication. After capturing a new sound segment, the controller ensures FFT Core readiness and triggers a hand-shaking protocol for data transfer.

The frequency-domain convolution operation initiates after computing the FFT values of the input segment and storing them in the single-port 17.4 kB FFT RAM. This is crucial as the convolution is independently performed 40 times with 40 kernels. For enhanced processing speed, normalized frequency-domain kernel values are stored in the dual-port Frequency-Domain (F-D) Kernel ROM, offering a storage capacity of 693 kB. The FFT RAM and F-D Kernel ROM values are sent to the Complex Multiplier for the subsequent convolution, where FFT values of the input segment are multiplied by stored FFT values of the kernels. The final convolution result is obtained by performing an inverse FFT (IFFT) on the multiplication results.

The Complex Multiplier executes 34×34 complex multiplications followed by an addition operation. The Complex Multiplier is designed using the Xilinx Complex Multiplier IP core. This IP core uses cascaded DSP slices for improved performance. It takes 10 clock cycles to compute the output and the results are then rounded to 34 bits. The implementation utilizes Xilinx FFT IP Core to calculate both forward (FFT) and inverse (IFFT) Fourier transforms. The Xilinx FFT Core can be configured in real time to calculate an N-point forward or inverse FFT, where $N$ can be $2^m$, and $m$ ranges from 3 to 16 [27]. The size of a full-resolution convolution (i.e. N, the size of the transform) output is given by $S + L - 1$, where $S$ denotes the number of samples in a segment and $L$ represents the number of samples in a kernel. Increasing the size of the transform results in increased hardware resource usage, while decreasing the transform size can impair encoding efficiency. As each kernel comprises $L = 1353$ samples (coefficients or taps), FFT Cores are designed with a size of $N = 2048$ samples ($m = 11$), which enables real-time processing while ensuring practical hardware utilization. The chosen transform size results in a segment size of $S = 696$ samples (equivalent to 43.5 ms).

The Xilinx FFT IP Core provides various architectural designs that offer different trade-offs between processing speed and hardware resource utilization. Among these, the Burst I/O architectures (also known as Radix-X) perform iterative computations for the Discrete Fourier Transform (DFT) on input data arrays. In the Radix-2 architecture, the algorithm handles two complex numbers from the input memory and produces the output in the complex number format. Similarly, the Radix-4 architecture processes four complex numbers in each iteration. On the other hand, Xilinx provides a Pipelined Streaming I/O architecture, optimizing performance by arranging multiple Radix-2 processing engines in a pipeline to ensure continuous data processing [27]. Given that the FFT of the input signal is computed only once in the spike generation process, the Radix-4 architecture is preferred for its optimal balance between processing speed and resource efficiency, even though it exhibits a slight performance decrease compared to the Pipelined architecture [27]. As the maximum data sample precision offered by Xilinx FFT IP Cores is 34-bit, the remaining processing blocks of Spiketrum are implemented with a resolution of 34 bits. On the other hand, the IFFT core is implemented using the Pipelined architecture to enable faster processing, given that the IFFT needs to be performed 40 times (after multiplying FFT of the input signal with all kernels).

The Code Generator continuously examines the output from the IFFT core and locates the maximum intensity $s_i$ following the completion of convolution operations across all kernels. When the IFFT operation concludes, it identifies the maximum convolution intensity $s_i$ along with its coordinates - kernel index $m_i$ and time shift $\tau_i$, using a 34-bit comparator that sequentially compares the outputs and find the highest convolution intensity. Upon the completion of convolution, the Code Generator dispatches the generated code $(m_i, \tau_i, s_i)$ to the Residual Computing and the Spike Generator processes (Intensity-to-Place Coding). After eliminating the detected code from the original signal using Residual Computing unit, the residue $x(t)_{new}$ is stored on the Signal RAM to extract the next dominant feature of the currently processed segment. The process of identifying subsequent matching kernels is repeated for a predetermined number of iterations per segment ($k$), resulting in a code set, represented as:

$$S = [(m_1, \tau_1, s_1), (m_2, \tau_2, s_2), \dots, (m_k, \tau_k, s_k)] \quad (2)$$

Every code $(m_i, \tau_i, s_i)$ within the code set $S$ represents a small sound waveform occurring at the temporal location $\tau_i$ with the shape of kernel $m_i$ and amplitude $s_i$. The error $\epsilon(t)$ decreases as the number of generated codes, $k$, increases. The original waveform can be reconstructed by linearly superposing all small waveforms represented by the code set $S$. The amount of information loss in the encoding process depends on the nature of the selected kernel set as well as the number of generated codes, $k$.

**Feedback Unit:** To illustrate the application of feedback in the Spiketrum algorithm, this implementation incorporates a simple stopping mechanism. This feature enables the encoder and other units to terminate the process when the convolution intensities of the last processed data fall below a predefined threshold. It's important to note that while we demonstrate this particular feedback mechanism, the Spiketrum algorithm is versatile enough to accommodate various feedback mechanisms. Here, we present this as an example, but it can be easily substituted with a different approach.

The Feedback module examines the latest generated code and instructs the Controller to either start a new code generation process or end the encoding procedure for a segment. By ignoring less significant features, this method minimizes resource usage and power consumption, fostering a more cost-effective and streamlined encoding process. This Feedback module is realized using a single comparator that compares the intensity of the generated code $s_i$, with the predefined threshold. In the results section, we demonstrate the effects of introducing feedback projections and compare the power consumption and accuracy performance of this method with the case where no feedback projections are used. This comparison is performed using a sound classification task, providing a practical gauge of the algorithm's performance in real-world scenarios.

*B. Residual Computing*

Spiketrum iteratively generates consecutive codes that encapsulate Gammatone patterns and features of the incoming auditory signal, which prioritizes the encoding of dominant features, while less significant features are processed subsequently, reflecting their relative importance. This is accomplished by eliminating the most recently encoded kernel component from the input segment after each feature extraction, before using the segment to generate a new code. The process of eliminating the kernel component from the currently processed segment is facilitated by the Residual Computing unit. This unit is comprised of three subunits: the Shifter, Multiplier, and Subtractor, as illustrated in Fig. 1-a.

Firstly, the time-domain (T-D) values of the corresponding kernel, denoted as $\emptyset_{m_i}$, are loaded from the T-D Kernel ROM and transferred to the Shifter unit. The T-D Kernel ROM is a single-port ROM that is equipped with a storage capacity of 346 kB. The Shifter unit facilitates segment shifting using a RAM-based approach, where a segment size of 2048 allows a maximum time shift value $\tau_i$ of ±1024. The RAM's starting writing address dynamically adjusts based on $\tau_i$, ensuring the preservation of the original kernel during read operations. Positive time shifts initiate a right-shift operation, while negative time shifts initiate a left-shift operation. Following each read operation, the RAM resets itself in preparation for subsequent shifts. With a storage capacity of 13 kB, the Shifter RAM effectively manages these operations, guaranteeing a seamless and efficient shifting process. Secondly, the shifted kernel is forwarded to a Multiplier unit, where it is scaled by the intensity of the captured code $s_i$. Real (non-complex) multiplication is performed using a single DSP48E1 Slice, which is known for its low-power consumption and ability to handle 25x18 two's-complement multiplications [28]. To ensure proper timing, both the input and output buses of the Multiplier are pipelined at three stages. Finally, the resulting output from the Multiplier is then sent to the Subtractor unit. Here, the shifted scaled kernel $s_i \emptyset_{m_i}(t - \tau_i)$ is subtracted from the current input segment $x(t)$ to obtain the updated segment $x(t)_{new}$, and it is written back to the Signal RAM. This segment is processed to generate a new code representing the next significant feature. The Subtractor unit is implemented using logic fabric on the FPGA, as the timing requirements for this simple operation can be easily met.

### C. Intensity-to-Place Coding

Upon the completion of the Feature Extraction process, the Intensity-to-Place (ITP) coding scheme maps the produced codes to their corresponding output channels (Fig. 1-a). The ITP scheme retrieves the generated code $(m_i, \tau_i, s_i)$ from the Code Generator and selects the channel intensity $C_i$ that best matches the code's intensity, $s_i$. Employing three subtractors, the system calculates the differences between the three channel intensities $(C_i)$ and the code intensity $(s_i)$. The closest channel intensity is then identified by assessing the smallest value among the outputs of the three subtractors, facilitated by the use of two comparators. Once this closest channel intensity has been determined, the Delay process, which is implemented by a counter, waits for a time delay equal to the time shift $\tau_i$ and then sends a spike at the selected output channel.

## IV. EXPERIMENTAL RESULTS

### A. Characterization of the Cochlea

We analysed the cochlea's behaviour to characterize it comprehensively. Fig 2-a demonstrates a clear one-to-one transformation between the input and output of the cochlea when using fundamental Gammatone kernel waveforms as the input. The waveforms of a subset of Gammatone kernels used are presented in Fig 2-d. The second characterization, depicted in Fig 2-b, involves exposing the cochlea's input to a sinusoidal sine waveform. The frequency logarithmically sweeps from 20 Hz to 8 kHz over a period of five seconds. This input, sampled at 16 kHz, initiates feature extraction and spike coding processes in our hardware prototype at a maximum speed of 200 MHz via USB-3. As the test input signal undergoes a logarithmic sweep alongside the implemented kernels, a distinct linear correlation becomes evident across the output channels. To enhance clarity, we standardize the output spike rate to 1 sps, arranging channels to portray a straightforward relationship between output channels and the frequency of the input signal. The raster plot depicted in Fig. 2-a and -b illustrates this correlation, with a more pronounced effect observed at higher frequencies. With a constant input signal amplitude, each processed segment activates a single channel per kernel. Fig. 2-e explores the encoding process at various spike rates (sps) for an input signal containing three spoken words ("One," "Two," and "Three" from the Google Speech Commands dataset). Higher spike rates, like 256 sps, capture finer features, emphasizing the need for careful spike rate selection tailored to specific applications. Notably, a spike rate of 1024 sps illustrates the risk of over-encoding, causing features to vanish. Our study highlighted the significant impact of spike rates on classification accuracy across various classifiers [3].

### B. Power Consumption and Resource Utilization

When processing in real-time at a frequency of 200 MHz, the device consumes a total power of 1.38 W, which includes both static and dynamic power. Fig. 2-c depicts a breakdown of dynamic power consumption, which constitutes 89% of the total power drawn when operating at the maximum speed. Power and hardware utilization statistics are estimated through Xilinx Vivado tools [29].

Undoubtedly, BRAMs and DSP slices emerge as pivotal elements in our implementation, collectively claiming over 50% of the power consumption. However, our paramount objective in this implementation was to delicately balancing power/resource utilization with encoding throughput (sps). Tailoring our approach to the specific requirements of diverse target applications, resource consumption can be further mitigated by opting for fabric logic over DSP slices or by locally generating FFT kernels on the board. This strategic adjustment yields a consequential reduction, approximately two-thirds, in BRAM utilization. Nevertheless, these optimizations come with a trade-off in terms of performance.

### C. Sound Classification Application

**Passive Spiketrum (with no feedback)**: In our prior research [3], we conducted an extensive benchmarking study of Spiketrum hardware and its software counterpart, comparing them to state-of-the-art biologically-inspired encoders (Spectrogram [30], Lauscher [4], and MAP [31]). Our evaluations encompassed various critical criteria, including classification accuracy, training speed, and sparsity, when utilizing encoder outputs for pattern recognition and classification tasks with both spiking classifiers, including Recurrent Spiking Neural Networks (RSNNs) and adaptive RSNN (aRSNNs), and non-spiking classifiers, including Convolutional Neural Networks (CNNs) and Long- and Short-Term Memory (LSTM). Additionally, we investigated encoded output entropy, hardware resource utilization, and power consumption in the hardware version of the encoders. The results of our study demonstrated Spiketrum's remarkable superiority across most benchmarking aspects, positioning it as a promising choice for diverse applications. Notably, Spiketrum efficiently utilized hardware resources, achieving high classification accuracy with relatively low power consumption. Furthermore, our research underscored the significant potential of encoders in spike-based processing, offering an avenue to enhance the efficiency and performance of neural computing systems. Table I presents a summary of the classification results of Spiketrum for different spike rates.

Table II, featuring classification accuracy, training time, information entropy and gain, and sparsity for Spiketrum with other benchmarked encoders. The evaluation of average classification accuracy spans four unique applications related to sound classification. The training duration serves as an indicator of the epochs necessary to achieve optimal accuracy. Furthermore, sparsity, which measures the ratio of active channels in spike-based encoding, exhibits variability among encoders, influencing both classification accuracy and spike count per sample. Notably, Spiketrum distinguishes

itself with commendable accuracy and a low sparsity, showcasing efficiency in computational load, energy consumption, and resource utilization within spike-based systems.

**Active Spiketrum:** In addition to the previously mentioned classification performances of the passive Spiketrum mentioned earlier, we now showcase the application of an active Spiketrum (with feedback) for sound classification tasks using an Artificial Neural Network (ANN) under varying noise levels. Our demonstration includes a straightforward feedback mechanism that assesses the convolution intensity of the latest code. Subsequently, it triggers a stop decision if the convolution intensity falls below a predetermined threshold. Our demonstration includes a straightforward feedback mechanism that assesses the convolution intensity of the most recent code and triggers a stop decision by comparing it with a predetermined threshold. The threshold value can be determined using a trial-and-error approach, allowing one to optimize the required balance between accuracy and power savings.

Here both the active and passive hardware implementations of Spiketrum are used to extract features from five different musical instrument classes in the MedleyDB71 dataset [32]: Drum Set, Flute, Piano, Trumpet, and Violin. Each class comprises 160 sound waveforms, with melodies and styles varying across a duration of 1.5 seconds each, sampled at a frequency of 16 kHz. The dataset is split equally between training and testing data subsets.

TABLE I. Summary of Spiketrum hardware's Encoding and Their Classification Performances

| Spike rate [sps] | Classifiers | Best Classification Accuracy [%] | Training Latency [no. epochs] | Entropy $\times 10^4$ [bit] |
|---|---|---|---|---|
| 64 | CNN | 96.94 | 30 | 2.46 |
| | LSTM | 96.66 | 45 | |
| | RSNN | 86.72 | 57 | |
| | aRSNN | 93.97 | 53 | |
| 128 | CNN | 98.79 | 54 | 2.85 |
| | **LSTM** | **98.98** | **13** | |
| | RSNN | 87.50 | 29 | |
| | aRSNN | 95.24 | 52 | |
| 256 | CNN | 99.44 | 51 | 3.07 |
| | LSTM | 99.44 | 57 | |
| | RSNN | 90.63 | 36 | |
| | aRSNN | 95.28 | 40 | |

TABLE II: Encoder Performance Across Classifiers and Experiments

| Encoder | Average classification accuracy [%] | Average training time [no. epochs] | Gain $\times 10^{-1}$ | Entropy $\times 10^4$ | Sparsity [%] |
|---|---|---|---|---|---|
| Spiketrum | 96 | 29 | 0.40 | 2.54 | 7 |
| Lauscher [4] | 94 | 48 | 5.39 | 14.58 | 10.2 |
| Spectrogram [30] | 88 | 52 | 9.98 | 1.23 | 4.5 |
| MAP [31] | 97 | 52 | 6.94 | 1.49 | 10 |

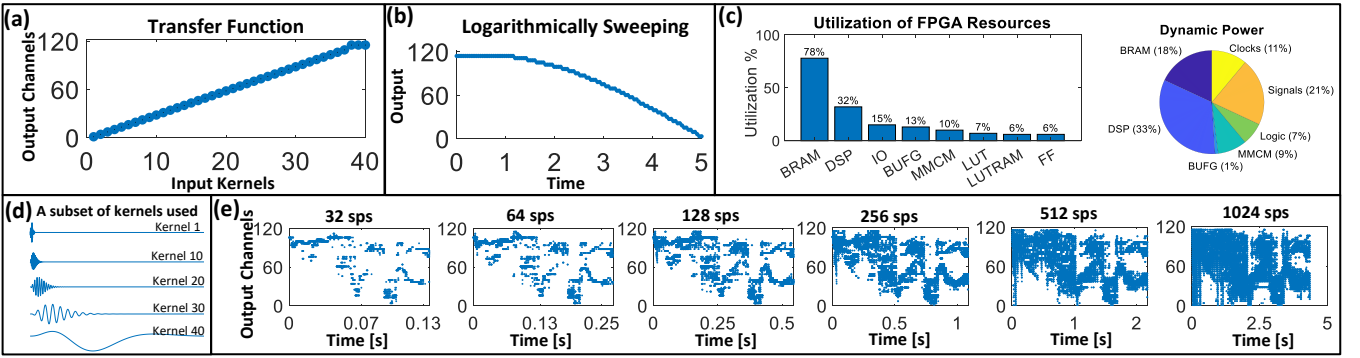

**Fig. 2**. Characteristic of the Spiketrum Hardware and examples of encoded spikes at different spike rates: **(a)** shows the transfer function of the Spiketrum, generated by feeding the audio signal constructed through sweeping the 40 kernels as audio input to the Spiketrum. **(b)** shows the output spikes recorded from the 120 output channels of the cochlea in response to an input sine wave with logarithmically increasing frequencies from 20 Hz to 8 kHz, maintaining a fixed amplitude of 1 $V_{pp}$. **(c)** The power and resource utilization breakdown of the cochlea's hardware implementation is provided. At 200 MHz, the cochlea consumes 1.38 W (dynamic and static), with dynamic power constituting 89% of the total. The chart illustrates the allocation of FPGA resources, including Block RAM tiles (BRAM), Digital Signal Processing slices (DSP), Input/output (IO), Global Clock Buffer (BUFG), Mixed-Mode Clock Manager Model (MMCM), Lookup tables (LUT), LUT RAMs, and Flip-Flops (FF). **(d)** A subset of implemented Gammatone kernels. **(e)** Spiketrum response corresponds to various spike rates when a three-spoken-words ("One," "Two," and "Three") taken from the GSC2 dataset [33] is used as the input signal.

To evaluate the performance of these implementations in noise environments, varying levels of noise, indicated by the signal-to-noise ratio (SNR) dB levels, are introduced into the dataset. This noise is represented by a background speech babble, featuring a 100-person conversation in a canteen, sampled at a rate of 19.98 kHz [34]. The input waveforms are subject to real-time encoding through the hardware setups, operating at a spike rate of 50 sps. In the active implementation, the generated spikes per segment range between a minimum of 1 sps and a maximum of 50 sps for the different samples and the introduced noise levels. The 50 sps spike rate translates to 1.15k spike per second. Afterward, the generated spike outputs are documented and introduced to a Deep Neural Network (DNN) for the purpose of sound classification. Developed using MATLAB's Deep Learning Toolbox, the DNN features a sequence input layer (3 nodes), followed by an LSTM layer, fully-connected layer, and SoftMax layer (each with 50 nodes), optimizing learning and memory retention. The final classification output layer comprises 5 nodes, aligning with the model's five distinct classes. This design facilitates efficient information processing from input to output. The training of this network was carried out using the cross-entropy loss function in combination with the ADAM optimizer [35]. The network underwent a training process spanning 30 epochs with a batch size set at 128 and a learning rate of 0.001, to reach the desired balance between speed and accuracy between the available solutions. The dataset was partitioned into training data, constituting 80% of the dataset, and test data, comprising the remaining 20%. The supervised training was executed using a single CPU.

In Fig. 3-a, DNN test classification results are presented using hardware-generated codes (before Intensity-to-Spike Coding) to distinguish five distinct musical instruments, with varying levels of noise added to the dataset. As expected, the classification accuracy of the DNN improves as the Signal-to-Noise Ratio (SNR) of the input signals increases. This improvement is observed for both implementations. However, the active implementation, which employs a predetermined fixed threshold, achieves greater power savings but results in a noticeable decrease in accuracy for the clean dataset, as illustrated in Fig. 3-b.

Fig. 3-b presents a comparative analysis of accuracy and reduction in power consumption for the two implementations, offering insights into their relative performance. Percentage differences highlight enhancements with a positive difference and deteriorations with a negative difference in the active implementation compared to the passive one. Power-saving results are directly calculated from the reduced number of generated spikes. It's noteworthy that varying threshold levels can influence the balance between accuracy and power savings, enabling users to configure the predefined threshold based on their specific application requirements.

## V. DISCUSSION

While Spiketrum can be efficiently implemented in advanced Very Large-Scale Integration (VLSI) technologies for enhanced energy efficiency and speed, our current focus is on validating the concept through FPGA-based implementation. The FPGA-based approach serves to confirm the feasibility in hardware and provides a versatile, easily configurable design platform for diverse applications. This adaptability empowers us to tailor and optimize the design to meet distinct use-cases and performance benchmarks. In the transformation of the Spiketrum algorithm into a tangible hardware entity, we have adopted a strategic approach involving various optimization goals and techniques. Our design intentionally incorporates Digital Signal Processing (DSP) slices and Look-Up Tables (LUTs), complemented by multi-stage pipelining, ensuring real-time processing within realistic hardware utilization constraints. For a comprehensive evaluation, Table III provides a detailed comparison of our proposed implementation with other cochlea implementations from the existing literature. This includes comparisons with implementations proposed by Ying Xu [5], Yang et al [36], Liu et al [37] and Fragnière [38]. Spiketrum excels in preserving spatiotemporal information with minimal encoding error, offering a broad dynamic input range and numerous output channels. This makes it an outstanding general-purpose neuromorphic encoder, ideal for spike coding in intelligent machines, communication systems, and data compression. Through benchmarking against state-of-the-art encoders, Spiketrum demonstrates superior performance in varying classification applications, showcasing its efficiency in hardware resource utilization with low power consumption and high classification accuracy.

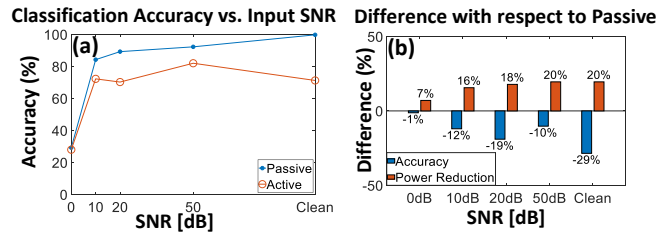

**Fig. 3**: DNN classification results using the passive and the active Spiketrum implementations. **(a)** The DNN test classification results of the hardware-generated codes. Here, five different musical instrument classes from MedleyDB68 dataset were encoded at a spike rate of 50 sps (or less for the active implementation). Each class contains 160 1.5-second-long sound waveforms sampled at 16 kHz. **(b)** For different input noise levels, the changes in accuracy and power consumption for the active feedback implementation are compared to those of the passive implementation, expressed as a percentage of the passive implementation, calculated as ((Active - Passive) / Passive * 100). The stop threshold is set to 0.01.

TABLE III. A Comparison with Existing Electronic Cochleae

| Imp.<br>Specs. | This work (2023) | [5] (2018) | [36] (2016) | [37] (2010) | [38] (2005) |
|---|---|---|---|---|---|
| | FPGA-based | | VLSI-based | | |
| Process Technology | 28 nm | 28 nm | 0.18 μm | 0.35 μm | 0.5 μm |
| Architecture based | Xilinx Artix-7 | Altera Cyclone-5 | CMOS | CMOS | CMOS |
| Filter Type | Parallel | Cascade | Parallel | Cascade | Parallel |
| Channel Number | 120 | 70 | 64 | 64 | 100 |
| Input Range (dB) | 110 (@ 1kHz) | 70 | 73 | 36 | 50 |
| Frequency Range (Hz) | 20-8,000 | Up to 22,050 | 8,000-20,000 | 50-20,000 | 200-20,000 |
| Power Supply (V) | 1.8 | 1.1 | 0.5 | - | 3.3 |
| Power Consumption (mW) | 1,382* | 1,260 | 0.055 | 59 | 1.7 |

*: estimated by Xilinx Vivadio Tools [29].

Furthermore, the optimization of Spiketrum, which introduces a straightforward stopping mechanism, serves as a demonstration of the delicate balance achieved between computational load, energy consumption, and accuracy. While this optimization yields tangible benefits, there is merit in exploring a more intricate and dynamic approach. This advanced approach could incorporate neural feedback signals from higher cortical regions, such as thalamocortical projections. By doing so, the system may achieve enhanced tunability and improved noise-cancelling capabilities, thereby extending its overall adaptability. This heightened adaptability holds promise in addressing more intricate challenges, such as the speaker identification task or the cocktail party problem, where elevated levels of noise or complex interference patterns pose significant hurdles.

## VI. CONCLUSION

In this work, we have effectively designed and constructed an FPGA-based prototype of a spike-encoding algorithm, thereby showcasing a neuromorphic cochlea capable of real-time operation. This innovative cochlea demonstrates proficiency in translating audio signals into spike trains, which can be easily integrated into larger neuromorphic systems, thereby opening the door for a wide range of intelligent machine applications. Spiketrum successfully demonstrated efficient resource utilization by employing both Artificial Neural Networks (ANN) and Spiking Neural Networks (SNN), achieving high accuracy with minimal power consumption. There is an ongoing plan to validate the implementation on a broader scale. This would potentially involve multiple diverse datasets and other sensory signals, including but not limited to Electrocardiogram (ECG) and Electromyogram (EMG). Through this expansion, we hope to demonstrate the versatility and robustness of our solution in dealing with various sensory data types.